\documentclass[orivec,runningheads]{llncs}
\usepackage[utf8]{inputenc}
\usepackage[T1]{fontenc}
\usepackage{microtype}
\usepackage{booktabs}
\usepackage{graphicx}
\usepackage{url}
\usepackage[numbers,sectionbib]{natbib}
\usepackage{listings}
\usepackage{amsthm}
\usepackage{amsmath}
\usepackage{amssymb}
\usepackage{xcolor}
\usepackage{array}
\usepackage[ruled]{algorithm2e}
\usepackage{booktabs}
\usepackage{mathpartir}
\usepackage{stmaryrd}

\usepackage[pdftex,pdfpagelabels,bookmarks,hyperindex,hyperfigures]{hyperref}
\usepackage[tt=false, type1=true]{libertine} 
\usepackage[varqu]{zi4} 

\let\llncssubparagraph\subparagraph
\let\subparagraph\paragraph
\usepackage{titlesec}
\let\subparagraph\llncssubparagraph

\titlespacing*{\subsection}{0pt}{1.0\baselineskip}{0.5\baselineskip}

\makeatletter
\theoremstyle{plain}
\newtheorem*{rep@theorem}{\rep@title}
\newcommand{\newreptheorem}[2]{%
\newenvironment{rep#1}[1]{%
 \def\rep@title{#2 \ref{##1}}%
 \begin{rep@theorem}}%
 {\end{rep@theorem}}}
\makeatother

\newreptheorem{theorem}{Theorem}
\newreptheorem{lemma}{Lemma}

\newcommand{\set}[1]{{\{{#1}\}}}

\newcommand{\tup}[1]{{\left\langle{#1}\right\rangle}}
\newcommand{\denot}[1]{{\llbracket{#1}\rrbracket}}

\newcommand{\pto}{\rightharpoonup}

\newcommand{\true}{\top}
\newcommand{\false}{\bot}

\DeclareMathOperator{\dom}{dom}

\newcommand{\code}[1]{\textnormal{\texttt{#1}}}

\newcommand{\civl}{\textsc{civl}}

\lstdefinelanguage{base}{
  moredelim=**[is][\color{red}]{|}{|},
  moredelim=**[is][\color{blue}]{@}{@},
  moredelim=**[is][\color{green!60!black}]{^}{^},
  basicstyle=\footnotesize\ttfamily,
  morekeywords={procedure, atomic, of, true, false, return, while, if, then, else, vis, lin, getModLin, absolute, monotonic, var, int, array},
  morecomment=*[s][\color{green!60!black}]{/*}{*/},
  morecomment=*[l][\color{green!60!black}]{//},
  columns=flexible,
  mathescape=true
}

\lstdefinelanguage{boogie}{
  morekeywords={acc, struct,if,else,returns,procedure,requires,ensures,:=,var,
    new,old,free,implicit,modifies,call,locals,assume,assert,choose,havoc,ghost,
    predicate,function,invariant,while, return,atomic, split, sync,
    mark, unmark, define, datatype, forall, int},
  basicstyle=\footnotesize\ttfamily,
  xleftmargin=2em,
  escapeinside={@}{@},
  columns=flexible,
  morecomment=*[s][\color{green!60!black}]{/*}{*/},
  morecomment=*[l][\color{green!60!black}]{//},
  mathescape=true,
}

\begin{document}
  \pagestyle{plain}
  \bibliographystyle{splncs04}
  \renewcommand{\bibname}{References}

  \title{Verifying Visibility-Based Weak Consistency}

  \author{Siddharth Krishna\inst{1} \and Michael Emmi\inst{2} \and Constantin Enea\inst{3} \and Dejan Jovanović\inst{2}}

  \institute{New York University, USA, \email{siddharth@cs.nyu.edu} \and
  SRI International, New York, NY, USA, \email{michael.emmi@gmail.com}, \email{dejan.jovanovic@sri.com} \and
  Université de Paris, IRIF, CNRS, F-75013 Paris, France, \email{cenea@irif.fr}}

  \maketitle

  \begin{abstract}

    Multithreaded programs generally leverage efficient and thread-safe \emph{concurrent objects} like sets, key-value maps, and queues. While some concurrent-object operations are designed to behave atomically, each witnessing the atomic effects of predecessors in a linearization order, others forego such strong consistency to avoid complex control and synchronization bottlenecks. For example, contains (value) methods of key-value maps may iterate through key-value entries without blocking concurrent updates, to avoid unwanted performance bottlenecks, and consequently overlook the effects of some linearization-order predecessors. While such \emph{weakly-consistent} operations may not be atomic, they still offer guarantees, e.g.,~only observing values that have been present.

    In this work we develop a methodology for proving that concurrent object implementations adhere to weak-consistency specifications. In particular, we consider (forward) simulation-based proofs of implementations against \emph{relaxed-visibility specifications}, which allow designated operations to overlook some of their linearization-order predecessors, i.e.,~behaving as if they never occurred. Besides annotating implementation code to identify \emph{linearization points}, i.e.,~points at which operations’ logical effects occur, we also annotate code to identify \emph{visible operations}, i.e.,~operations whose effects are observed; in practice this annotation can be done automatically by tracking the writers to each accessed memory location. We formalize our methodology over a general notion of transition systems, agnostic to any particular programming language or memory model, and demonstrate its application, using automated theorem provers, by verifying models of Java concurrent object implementations.

  \end{abstract}


\section{Introduction}
\label{sec:introduction}

Programming efficient multithreaded programs generally involves carefully organizing shared memory accesses to facilitate inter-thread communication while avoiding synchronization bottlenecks. Modern software platforms like Java include reusable abstractions which encapsulate low-level shared memory accesses and synchronization into familiar high-level abstract data types (ADTs). These so-called \emph{concurrent objects} 
typically include mutual-exclusion primitives like locks, numeric data types like atomic integers, as well as collections like sets, key-value maps, and queues; Java’s standard-edition platform contains many implementations of each. Such objects typically provide strong consistency guarantees like \emph{linearizability}~\cite{DBLP:journals/toplas/HerlihyW90}, ensuring that each operation appears to happen atomically, witnessing the atomic effects of predecessors according to some linearization order among concurrently-executing operations.

While such strong consistency guarantees are ideal for logical reasoning about programs which use concurrent objects, these guarantees are too strong for many operations, since they preclude simple and/or efficient implementation — over half of Java’s concurrent collection methods forego atomicity for \emph{weak-consistency}~\cite{DBLP:journals/pacmpl/EmmiE19}. On the one hand, basic operations like the get and put methods of key-value maps typically admit relatively-simple atomic implementations, since their behaviors essentially depend upon individual memory cells, e.g.,~where the relevant key-value mapping is stored. On the other hand, making aggregate operations like size and contains (value) atomic would impose synchronization bottlenecks, or otherwise-complex control structures, since their atomic behavior depends simultaneously upon the values stored across many memory cells. Interestingly, such implementations are not linearizable even when their underlying memory operations are sequentially consistent, e.g.,~as is generally the case with Java’s concurrent collections, whose memory accesses are data-race free.

For instance, the contains (value) method of Java’s concurrent hash map iterates through key-value entries without blocking concurrent updates in order to avoid unreasonable performance bottlenecks. Consequently, in a given execution, a contains-value-$v$ operation~$o_1$ will overlook operation~$o_2$’s concurrent insertion of $k_1 \mapsto v$ for a key $k_1$ it has already traversed. This oversight makes it possible for $o_1$ to conclude that value $v$ is not present, and can only be explained by $o_1$ being linearized before $o_2$. In the case that operation~$o_3$ removes $k_2 \mapsto v$ concurrently before $o_1$ reaches key $k_2$, but only after $o_2$ completes, then atomicity is violated since in every possible linearization, either mapping $k_2 \mapsto v$ or $k_1 \mapsto v$ is always present. Nevertheless, such weakly-consistent operations still offer guarantees, e.g.,~that values never present are never observed, and initially-present values not removed are observed.

In this work we develop a methodology for proving that concurrent-object implementations adhere to the guarantees prescribed by their weak-consistency specifications. The key salient aspects of our approach are the lifting of existing sequential ADT specifications via \emph{visibility relaxation}~\cite{DBLP:journals/pacmpl/EmmiE19}, and the harnessing of simple and mechanizable reasoning based on \emph{forward simulation}~\cite{DBLP:journals/iandc/LynchV95} by relaxed-visibility ADTs. Effectively, our methodology extends the predominant forward-simulation based linearizability-proof methodology to concurrent objects with weakly-consistent operations, and enables automation for proving weak-consistency guarantees.

To enable the harnessing of existing sequential ADT specifications, we adopt the recent methodology of \emph{visibility relaxation}~\cite{DBLP:journals/pacmpl/EmmiE19}. As in linearizability~\cite{DBLP:journals/toplas/HerlihyW90}, the return value of each operation is dictated by the atomic effects of its predecessors in some (i.e.,~existentially quantified) linearization order. To allow consistency weakening, operations are allowed, to a certain extent, to overlook some of their linearization-order predecessors, behaving as if they had not occurred. Intuitively, this (also existentially quantified) \emph{visibility} captures the inability or unwillingness to atomically observe the values stored across many memory cells. To provide guarantees, the extent of visibility relaxation is bounded to varying degrees. Notably, the visibility of an \emph{absolute} operation must include all of its linearization-order predecessors, while the visibility of a \emph{monotonic} operation must include all happens-before predecessors, along with all operations visible to them. The majority of Java’s concurrent collection methods are absolute or monotonic~\cite{DBLP:journals/pacmpl/EmmiE19}. For instance, in the contains-value example described above, by considering that operation~$o_2$ is not visible to $o_1$, the conclusion that $v$ is not present can be justified by the linearization $o_2; o_3; o_1$, in which $o_1$ sees $o_3$’s removal of $k_2 \mapsto v$ yet not $o_2$’s insertion of $k_1 \mapsto v$. Ascribing the monotonic visibility to the contains-value method amounts to a guarantee that initially-present values are observed unless removed (i.e.,~concurrently).

While relaxed-visibility specifications provide a means to describing the guarantees provided by weakly-consistent concurrent-object operations, systematically establishing implementations’ adherence requires a strategy for demonstrating \emph{simulation}~\cite{DBLP:journals/iandc/LynchV95}, i.e.,~that each step of the implementation is simulated by some step of (an operational representation of) the specification. The crux of our contribution is thus threefold: first, to identify the relevant specification-level actions with which to relate implementation-level transitions; second, to identify implementation-level annotations relating transitions to specification-level actions; and third, to develop strategies for devising such annotations systematically. For instance, the existing methodology based on \emph{linearization points}~\cite{DBLP:journals/toplas/HerlihyW90} essentially amounts to annotating implementation-level transitions with the points at which its specification-level action, i.e.,~its atomic effect, occurs. Relaxed-visibility specifications require not only a witness for the existentially-quantified linearization order, but also an existentially-quantified visibility relation, and thus requires a second kind of annotation to signal operations’ visibilities. We propose a notion of \emph{visibility actions} which enable operations to assert their visibility of others, e.g.,~signaled for the writers of memory cells it has read.

The remainder of our approach amounts to devising a systematic means for constructing simulation proofs to enable automated verification. Essentially, we identify a strategy for systematically annotating implementations with visibility actions, given linearization-point annotations and visibility bounds (i.e.,~absolute or monotonic), and then encode the corresponding simulation check using an off-the-shelf verification tool. For the latter, we leverage \civl{}~\cite{DBLP:conf/cav/HawblitzelPQT15}, a language and verifier for Owicki-Gries style modular proofs of concurrent programs with arbitrarily-many threads. 
In principle, since our approach reduces simulation to safety verification, any safety verifier could be used, though \civl{} facilitates reasoning for multithreaded programs by capturing interference at arbitrary program points. Using \civl{}, we have verified monotonicity of the contains-value and size methods of Java’s concurrent hash-map and concurrent linked-queue, respectively — and absolute consistency of add and remove operations. Although our models are written in \civl{} and assume sequentially-consistent memory accesses, they capture the difficult aspects of weak-consistency in Java, including heap-based memory access; furthermore, our models are also sound with respect to Java’s weak memory model, since their actual Java implementations guarantee data-race freedom by accessing individual shared-memory cells with atomic operations via volatile variables and compare-and-swap instructions.

In summary, we present the first methodology for verifying weakly-consistent operations using sequential specifications and forward simulation. Contributions include:
\begin{itemize}

  \item the formalization of our methodology over a general notion of transition systems, agnostic to any particular programming language or memory model (§3);

  \item the application of our methodology to verifying a weakly-consistent contains-value method of a key-value map (§4); and

  \item a mechanization of our methodology used for verifying models of weakly-consistent Java methods using automated theorem provers (§5).

\end{itemize}
Aside from the outline above, this article summarizes an existing weak-consistency specification methodology via visibility relaxation (§2), summarizes related work (§6), and concludes (§7). Proofs to all theorems and lemmas are listed in Appendix~\ref{sec:proofs}.

\section{Weak Consistency}
\label{sec:consistency}

Our methodology for verifying weakly-consistent concurrent objects relies both on the precise characterization of weak consistency specifications, as well as a proof technique for establishing adherence to specifications. In this section we recall and outline a characterization called \emph{visibility relaxation}~\cite{DBLP:journals/pacmpl/EmmiE19}, an extension of sequential abstract data type (ADT) specifications in which the return values of some operations may not reflect the effects of previously-effectuated operations.

Notationally, in the remainder of this article, $\varepsilon$ denotes the empty sequence, $\emptyset$ denotes the empty set, $\_$ denotes an unused binding, and $\true$ and $\false$ denote the Boolean values true and false, respectively. We write $R(x)$ to denote the inclusion $x \in R$ of a tuple $x$ in the relation $R$; and $R[x \mapsto y]$ to denote the extension $R \cup \set{ xy }$ of $R$ to include $xy$; and $R \mid X$ to denote the projection $R \cap X^\ast$ of $R$ to set $X$; and $\overline{R}$ to denote the complement $\set{ x : x \notin R }$ of $R$; and $R(x)$ to denote the image $\set{ y : xy \in R }$ of $R$ on $x$; and $R^{-1}(y)$ to denote the pre-image $\set{ x : xy \in R }$ of $R$ on $y$; whether $R(x)$ refers to inclusion or an image will be clear from its context. Finally, we write $x_i$ to refer to the $i$th element of tuple $x = x_0 x_1 \ldots$.

\subsection{Weak-Visibility Specifications}

For a general notion of ADT specifications, we consider fixed sets $\mathbb{M}$ and $\mathbb{X}$ of method names and argument or return values, respectively. An \emph{operation label} $\lambda = \tup{m, x, y}$ is a method name $m \in \mathbb{M}$ along with argument and return values $x, y \in \mathbb{X}$. A \emph{read-only predicate} is a unary relation $R(\lambda)$ on operation labels, an \emph{operation sequence} $s = \lambda_0 \lambda_1 \ldots$ is a sequence of operation labels, and a \emph{sequential specification} $S = \set{ s_0, s_1, \ldots }$ is a set of operation sequences. We say that $R$ is \emph{compatible} with $S$ when $S$ is closed under deletion of read-only operations, i.e.,~$\lambda_0 \ldots \lambda_{j-1} \lambda_{j+1} \ldots \lambda_i \in S$ when $\lambda_0 \ldots \lambda_i \in S$ and $R(\lambda_j)$.

\begin{example}
  \label{ex:spec:seq}

  The \emph{key-value map} ADT sequential specification $S_{\mathrm{m}}$ is the prefix-closed set containing all sequences $\lambda_0 \ldots \lambda_i$ such that $\lambda_i$ is either:
  \begin{itemize}

    \item $\tup{ \text{put}, kv, b }$, and $b = \true$ if{f} some $\tup{ \text{rem}, k, \_ }$ follows any prior $\tup{ \text{put}, kv, \_ }$;

    \item $\tup{ \text{rem}, k, b }$, and $b = \true$ if{f} no other $\tup{ \text{rem}, k, \_ }$ follows some prior $\tup{ \text{put}, kv, \_ }$;

    \item $\tup{ \text{get}, k, v }$, and no $\tup{ \text{put}, kv', \_ }$ nor $\tup{ \text{rem}, k, \_ }$ follows some prior $\tup{ \text{put}, kv, \_ }$, and $v = \bot$ if no such $\tup{ \text{put}, kv, \_ }$ exists; or

    \item $\tup{ \text{has}, v, b }$, and $b = \true$ if{f} no prior $\tup{ \text{put}, kv', \_ }$ nor $\tup{ \text{rem}, k, \_ }$ follows prior $\tup{ \text{put}, kv, \_ }$.

  \end{itemize}
  The read-only predicate $R_{\mathrm{m}}$ holds for the following cases:
  \begin{align*}
    R_{\mathrm{m}}(\tup{ \text{put}, \_, b }) \text{ if } \lnot b
    \qquad
    R_{\mathrm{m}}(\tup{ \text{rem}, \_, b }) \text{ if } \lnot b
    \qquad
    R_{\mathrm{m}}(\tup{ \text{get}, \_, \_ })
    \qquad
    R_{\mathrm{m}}(\tup{ \text{has}, \_, \_ }) \text{.}
  \end{align*}
  This is a simplification of Java’s Map ADT, i.e.,~with fewer methods.\footnote{For brevity, we abbreviate Java’s remove and contains-value methods by rem and has.}

\end{example}

To derive weak specifications from sequential ones, we consider a set $\mathbb{V}$ of exactly two \emph{visibility} labels from prior work~\cite{DBLP:journals/pacmpl/EmmiE19}: \emph{absolute} and \emph{monotonic}.\footnote{Previous work refers to absolute visibility as \emph{complete}, and includes additional visibility labels.} Intuitively, absolute visibility requires operations to observe the effects of all of their linearization-order predecessors, while monotonic visibility allows them to ignore effects which have been ignored by happens-before (i.e.,~program- and synchronization-order) predecessors. A \emph{visibility annotation} $V : \mathbb{M} \to \mathbb{V}$ maps each method $m \in \mathbb{M}$ to a visibility $V(m) \in \mathbb{V}$.

\begin{definition}
  \label{def:specification}

  A \emph{weak-visibility specification} $W = \tup{S, R, V}$ is a sequential specification $S$ with a compatible read-only predicate $R$ and a visibility annotation $V$.

\end{definition}

\begin{example}
  \label{ex:spec:weak}

  The \emph{weakly-consistent contains-value map} $W_{\mathrm{m}} = \tup{ S_{\mathrm{m}}, R_{\mathrm{m}}, V_{\mathrm{m}} }$ annotates the key-value map ADT methods of $S_{\mathrm{m}}$ from Example~\ref{ex:spec:seq} with:
  \begin{align*}
    V_{\mathrm{m}}(\text{put}) =
    V_{\mathrm{m}}(\text{rem}) =
    V_{\mathrm{m}}(\text{get}) =
    \text{absolute,}
    \qquad
    V_{\mathrm{m}}(\text{has}) = \text{monotonic.}
  \end{align*}
  Java’s concurrent hash map appears to be consistent with this specification~\cite{DBLP:journals/pacmpl/EmmiE19}.

\end{example}

We ascribe semantics to specifications by characterizing the values returned by concurrent method invocations, given constraints on invocation order. In practice, the \emph{happens-before} order among invocations is determined by a \emph{program order}, i.e.,~among invocations of the same thread, and a \emph{synchronization order}, i.e.,~among invocations of distinct threads accessing the same atomic objects, e.g.,~locks. A \emph{history} $h = \tup{O, \mathit{inv}, \mathit{ret}, \mathit{hb}}$ is a set $O \subseteq \mathbb{N}$ of numeric operation identifiers, along with an invocation function $\mathit{inv}: O \to \mathbb{M} \times \mathbb{X}$ mapping operations to method names and argument values, a partial return function $\mathit{ret}: O \pto \mathbb{X}$ mapping operations to return values, and a (strict) partial happens-before relation $\mathit{hb} \subseteq O \times O$; the \emph{empty history} $h_{\emptyset}$ has $O = \mathit{inv} = \mathit{ret} = \mathit{hb} = \emptyset$. An operation $o \in O$ is \emph{complete} when $\mathit{ret}(o)$ is defined, and is otherwise \emph{incomplete}; then $h$ is \emph{complete} when each operation is. The \emph{label} of a complete operation $o$ with $\mathit{inv}(o) = \tup{m,x}$ and $\mathit{ret}(o) = y$ is $\tup{ m, x, y }$.

To relate operations’ return values in a given history back to sequential specifications, we consider certain sequencings of those operations. A \emph{linearization} of a history $h = \tup{O, \_, \_, \mathit{hb}}$ is a total order $\mathit{lin} \supseteq \mathit{hb}$ over $O$ which includes $\mathit{hb}$, and a \emph{visibility projection} $\mathit{vis}$ of $\mathit{lin}$ maps each operation $o \in O$ to a subset $\mathit{vis}(o) \subseteq \mathit{lin}^{-1}(o)$ of the operations preceding $o$ in $\mathit{lin}$; note that $\tup{o_1,o_2} \in \mathit{vis}$ means $o_1$ observes $o_2$. For a given read-only predicate $R$, we say $o$’s visibility is \emph{monotonic} when it includes every happens-before predecessor, and operation visible to a happens-before predecessor, which is not read-only,\footnote{For convenience we rephrase Emmi and Enea~\cite{DBLP:journals/pacmpl/EmmiE19}’s notion to ignore read-only predecessors.} i.e.,~$\mathit{vis}(o) \supseteq \left( \mathit{hb}^{-1}(o) \mathrel\cup \mathit{vis}(\mathit{hb}^{-1}(o)) \right) \mid \overline{R}$. We says $o$’s visibility is \emph{absolute} when $\mathit{vis}(o) = \mathit{lin}^{-1}(o)$, and $\mathit{vis}$ is itself \emph{absolute} when each $\mathit{vis}(o)$ is. An \emph{abstract execution} $e = \tup{h, \mathit{lin}, \mathit{vis}}$ is a history $h$ along with a linearization of $h$, and a visibility projection $\mathit{vis}$ of $\mathit{lin}$. An abstract execution is \emph{sequential} when $\mathit{hb}$ is total, \emph{complete} when $h$ is, and \emph{absolute} when $\mathit{vis}$ is.

\begin{example}
  \label{ex:execution}

  An abstract execution can be defined using the linearization\footnote{For readability, we list linearization sequences with operation labels in place of identifiers.}
  \begin{align*}
    \tup{ \text{put}, \tup{1,1}, \true }
    \ \tup{ \text{get}, 1, 1 }
    \ \tup{ \text{put}, \tup{0,1}, \true }
    \ \tup{ \text{put}, \tup{1,0}, \false }
    \ \tup{ \text{has}, 1, \false }
  \end{align*}
 along with a happens-before order that compared to linearization order, keeps $\tup{ \text{has}, 1, \false }$ unordered w.r.t. $\tup{ \text{put}, \tup{0,1}, \true }$ and $\tup{ \text{put}, \tup{1,0}, \false }$, and a visibility projection where the visibility of every put and get includes all the linearization predecessors and the visibility of $\tup{ \text{has}, 1, \false }$ consists of $\tup{ \text{put}, \tup{1,1}, \true }$ and $\tup{ \text{put}, \tup{1,0}, \false }$.

\end{example}

To determine the consistency of individual histories against weak-visibility specifications, we consider adherence of their corresponding abstract executions. Let $h = \tup{O, \mathit{inv}, \mathit{ret}, \mathit{hb}}$ be a history and $e = \tup{h, \mathit{lin}, \mathit{vis}}$ a complete abstract execution. Then $e$ is \emph{consistent} with a visibility annotation $V$ and read-only predicate $R$ if for each operation $o \in \dom(\mathit{lin})$ with $\mathit{inv}(o) = \tup{m, \_}$: $\mathit{vis}(o)$ is absolute or monotonic, respectively, according to $V(m)$. The \emph{labeling} $\lambda_0 \lambda_1 \ldots$ of a total order $o_0 \prec o_1 \prec \ldots$ of complete operations is the sequence of operation labels, i.e.,~$\lambda_i$ is the label of $o_i$. Then $e$ is \emph{consistent} with a sequential specification $S$ when the labeling\footnote{As is standard, adequate labelings of incomplete executions are obtained by completing each linearized yet pending operation with some arbitrarily-chosen return value~\cite{DBLP:journals/toplas/HerlihyW90}. It is sufficient that one of these completions be included in the sequential specification.} of $\mathit{lin} \mid (\mathit{vis}(o) \cup \set{ o })$ is included in $S$, for each operation $o \in \dom(\mathit{lin})$.\footnote{We consider a simplification from prior work~\cite{DBLP:journals/pacmpl/EmmiE19}: rather than allowing the observers of a given operation to pretend they see distinct return values, we suppose that all observers agree on return values. While this is more restrictive in principle, it is equivalent for the simple specifications studied in this article.} Finally, we say $e$ is \emph{consistent} with a weak-visibility specification $\tup{S,R,V}$ when it is consistent with $S$, $R$, and $V$.

\begin{example}

  The execution in Example~\ref{ex:execution} is consistent with the weakly-consistent contains-value map $W_{\text{m}}$ defined in Example~\ref{ex:spec:weak}.

\end{example}

\begin{remark}

  Consistency models suited for modern software platforms like Java are based on \emph{happens-before} relations which abstract away from \emph{real-time} execution order. Since happens-before, unlike real-time, is not necessarily an \emph{interval order}, the composition of linearizations of two distinct objects in the same execution may be cyclic, i.e.,~not linearizable. Recovering compositionality in this setting is orthogonal to our work of proving consistency against a given model, and is explored elsewhere~\cite{DBLP:conf/vmcai/DongolJRA18}.

\end{remark}

The \emph{abstract executions} $E(W)$ of a weak-visibility specification $W = \tup{S,R,V}$ include those complete, sequential, and absolute abstract executions derived from sequences of $S$, i.e.,~when $s = \lambda_0 \ldots \lambda_n \in S$ then each $e_s$ labels each $o_i$ by $\lambda_i$, and orders $\mathit{hb}(o_i, o_j)$ if{f} $i < j$. In addition, when $E(W)$ includes an abstract execution $\tup{h, \mathit{lin}, \mathit{vis}}$ with $h = \tup{O, \mathit{inv}, \mathit{ret}, \mathit{hb}}$, then $E(W)$ also includes any:
\begin{itemize}

  \item execution $\tup{h', \mathit{lin}, \mathit{vis}}$ such that $h' = \tup{O, \mathit{inv}, \mathit{ret}, \mathit{hb}'}$ and $\mathit{hb'} \subseteq \mathit{hb}$; and

  \item $W$-consistent execution $\tup{h', \mathit{lin}, \mathit{vis}'}$ with $h' = \tup{O, \mathit{inv}, \mathit{ret}', \mathit{hb}}$ and $\mathit{vis}' \subseteq \mathit{vis}$.

\end{itemize}
Note that while \emph{happens-before weakening} $\mathit{hb}' \subseteq \mathit{hb}$ always yields consistent executions, unguarded \emph{visibility weakening} $\mathit{vis}' \subseteq \mathit{vis}$ generally breaks consistency with visibility annotations and sequential specifications: visibilities can become non-monotonic, and return values can change when operations observe fewer operations’ effects.

\begin{lemma}
  \label{lem:executions}

  The abstract executions~$E(W)$ of a specification~$W$ are consistent with $W$.

\end{lemma}

\begin{example}
  \label{ex:executions}

  The abstract executions of $W_{\mathrm{m}}$ include the complete, sequential, and absolute abstract execution defined by the following happens-before order
  \begin{align*}
  \tup{ \text{put}, \tup{1,1}, \true }
  \ \tup{ \text{get}, 1, 1 }
  \ \tup{ \text{put}, \tup{0,1}, \true }
  \ \tup{ \text{put}, \tup{1,0}, \false }
  \ \tup{ \text{has}, 1, \true }
  \end{align*}
  which implies that it also includes one in which just the happens-before order is modified such that $\tup{ \text{has}, 1, \true }$ becomes unordered w.r.t. $\tup{ \text{put}, \tup{0,1}, \true }$ and $\tup{ \text{put}, \tup{1,0}, \false }$. Since it includes the latter, it also includes the execution in Example~\ref{ex:execution} where the visibility of $\text{has}$ is weakened which also modifies its return value from $\true$ to $\false$.

\end{example}

\begin{definition}
  \label{def:histories}

  The \emph{histories} of a weak-visibility specification $W$ are the projections $H(W) = \set{ h : \tup{h, \_, \_} \in E(W) }$ of its abstract executions.

\end{definition}

\subsection{Consistency against Weak-Visibility Specifications}
\label{ssec:sim}

\newcommand{\vecaioverarrow}[1][]{\mathrel{%
  \raisebox{-2pt}{$\xrightarrow{\vec{a}_{#1}}$}}%
}

To define the consistency of implementations against specifications, we leverage a general model of computation to capture the behavior of typical concurrent systems, e.g.,~including multiprocess and multithreaded systems. A \emph{sequence-labeled transition system} $\tup{Q, A, q, \to}$ is a set $Q$ of states, along with a set $A$ of actions, initial state $q \in Q$ and transition relation $\mathbin{\to} \in Q \times A^\ast \times Q$. An \emph{execution} is an alternating sequence $\eta = q_0 \vec{a}_0 q_1 \vec{a}_1 \ldots q_n$ of states and action sequences starting with $q_0 = q$ such that $q_i \vecaioverarrow[i] q_{i+1}$ for each $0 \le i < n$. The \emph{trace} $\tau \in A^\ast$ of the execution $\eta$ is its projection $\vec{a}_0 \vec{a}_1 \ldots$ to individual actions.

To capture the histories admitted by a given implementation, we consider sequence-labeled transition systems (SLTSs) which expose actions corresponding to method call, return, and happens-before constraints. We refer to the actions call$(o,m,x)$, ret$(o,y)$, and hb$(o,o')$, for $o, o' \in \mathbb{N}$, $m \in \mathbb{M}$, and $x, y \in \mathbb{X}$, as \emph{the history actions}, and a \emph{history transition system} is an SLTS whose actions include the history actions. We say that an action over operation identifier $o$ is an \emph{$o$-action}, and assume that executions are \emph{well formed} in the sense that for a given operation identifier $o$: at most one call $o$-action occurs, at most one ret $o$-action occurs, and no ret nor hb $o$-actions occur prior to a call $o$-action. 
Furthermore, we assume call $o$-actions are enabled, so long as no prior call $o$-action has occurred. The \emph{history} of a trace $\tau$ is defined inductively by $f_{\mathrm{h}}(h_{\emptyset}, \tau)$, where $h_{\emptyset}$ is the empty history, and,
\begin{align*}
  \begin{array}{lcl}
    f_{\mathrm{h}}(h, \varepsilon) & = & h \\
    f_{\mathrm{h}}(h, a \tau) & = & f_{\mathrm{h}}(g_{\mathrm{h}}(h,a), \tau) \\
    f_{\mathrm{h}}(h, \tilde a \tau) & = & f_{\mathrm{h}}(h, \tau)
  \end{array}
  & &
  \begin{array}{lcl}
    g_{\mathrm{h}}(h, \textrm{call}(o,m,x))
    & =
    & \tup{O \cup \set{o}, \mathit{inv}[o \mapsto \tup{m,x}], \mathit{ret}, \mathit{hb}} \\
    g_{\mathrm{h}}(h, \textrm{ret}(o,y))
    & =
    & \tup{O, \mathit{inv}, \mathit{ret}[o \mapsto y], \mathit{hb}} \\
    g_{\mathrm{h}}(h, \textrm{hb}(o,o'))
    & =
    & \tup{O, \mathit{inv}, \mathit{ret}, \mathit{hb} \cup \tup{o,o'}}
  \end{array}
\end{align*}
where $h = \tup{O, \mathit{inv}, \mathit{ret}, \mathit{hb}}$, and $a$ is a call, ret, or hb action, and $\tilde a$ is not. An \emph{implementation} $I$ is a history transition system, and the \emph{histories} $H(I)$ of $I$ are those of its traces. Finally, we define consistency against specifications via history containment.

\begin{definition}
  \label{def:consistency}

  Implementation $I$ is \emph{consistent} with specification $W$ if{f} $H(I) \subseteq H(W)$.

\end{definition}


\section{Establishing Consistency with Forward Simulation}
\label{sec:simulation}

To obtain a consistency proof strategy, we more closely relate implementations to specifications via their admitted abstract executions. To capture the abstract executions admitted by a given implementation, we consider SLTSs which expose not only history-related actions, but also actions witnessing linearization and visibility. We refer to the actions lin$(o)$ and vis$(o,o')$ for $o, o' \in \mathbb{N}$, along with the history actions, as \emph{the abstract-execution actions}, and an \emph{abstract-execution transition system} (AETS) is an SLTS whose actions include the abstract-execution actions. Extending the corresponding notion from history transition systems, we assume that executions are \emph{well formed} in the sense that for a given operation identifier $o$: at most one lin $o$-action occurs, and no lin or vis $o$-actions occur prior to a call $o$-action. The \emph{abstract execution} of a trace $\tau$ is defined inductively by $f_{\mathrm{e}}(e_{\emptyset}, \tau)$, where $e_{\emptyset} = \tup{ h_{\emptyset}, \emptyset, \emptyset }$ is the empty execution, and,
\begin{align*}
  \begin{array}{lcl}
    f_{\mathrm{e}}(e, \varepsilon) & = & e \\
    f_{\mathrm{e}}(e, a \tau) & = & f_{\mathrm{e}}(g_{\mathrm{e}}(e,a), \tau) \\
    f_{\mathrm{e}}(e, \tilde a \tau) & = & f_{\mathrm{e}}(e, \tau)
  \end{array}
  & &
  \begin{array}{lcl}
    g_{\mathrm{e}}(e, \hat a)
    & =
    & \tup{ g_{\mathrm{h}}(h), \mathit{lin}, \mathit{vis}} \\
    g_{\mathrm{e}}(e, \textrm{lin}(o))
    & =
    & \tup{h, \mathit{lin} \cup \set{ \tup{o',o} : o' \in \mathit{lin} }, \mathit{vis}} \\
    g_{\mathrm{e}}(e, \textrm{vis}(o,o'))
    & =
    & \tup{h, \mathit{lin}, \mathit{vis} \cup \set{ \tup{ o, o' }}}
  \end{array}
\end{align*}
where $e = \tup{h, \mathit{lin}, \mathit{vis}}$, and $a$ is a call, ret, hb, lin, or vis action, $\tilde a$ is not, and $\hat a$ is a call, ret, or hb action. A \emph{witnessing implementation} $I$ is an abstract-execution transition system, and the \emph{abstract executions} $E(I)$ of $I$ are those of its traces.

We adopt forward simulation~\cite{DBLP:journals/iandc/LynchV95} for proving consistency against weak-visibility specifications. Formally, a \emph{simulation relation} from one system $\Sigma_1 = \tup{Q_1, A_1, \chi_1, \to_1}$ to another $\Sigma_2 = \tup{Q_2, A_2, \chi_2, \to_2}$ is a binary relation $R \subseteq Q_1 \times Q_2$ such that initial states are related, $R(\chi_1, \chi_2)$, and: for any pair of related states $R(q_1, q_2)$ and source-system transition $q_1 \vecaioverarrow[1]_1 q_1'$, there exists a target-system transition $q_2 \vecaioverarrow[2]_2 q_2'$ such that $R(q_1', q_2')$ and $(\vec{a}_1 \mid A_2) = (\vec{a}_2 \mid A_1)$. We say $\Sigma_2$ \emph{simulates} $\Sigma_1$ and write $\Sigma_1 \sqsubseteq \Sigma_2$ when a simulation relation from $\Sigma_1$ to $\Sigma_2$ exists.

We derive transition systems to model consistency specifications in simulation. The following lemma establishes the soundness and completeness of this substitution, and the subsequent theorem asserts the soundness of the simulation-based proof strategy.

\begin{definition}

  The \emph{transition system} $\denot{W}_\mathrm{s}$ of a weak-visibility specification $W$ is the AETS whose actions are the abstract execution actions, whose states are abstract executions, whose initial state is the empty execution, and whose transitions include $e_1 \vecaioverarrow e_2$ if{f} $f_{\mathrm{e}}(e_1, \vec{a}) = e_2$ and $e_2$ is consistent with $W$.

\end{definition}

\begin{lemma}
  \label{lem:histories}

  A weak-visibility spec. and its transition system have identical histories.

\end{lemma}

\begin{theorem}
  \label{thm:simulation}

  A witnessing implementation~$I$ is consistent with a weak-visibility specification~$W$ if the transition system $\denot{W}_\mathrm{s}$ of $W$ simulates $I$.

\end{theorem}

Our notion of simulation is in some sense \emph{complete} when the sequential specification $S$ of a weak-consistency specification $W = \tup{S, R, V}$ is \emph{return-value deterministic}, i.e.,~there is a single label $\tup{ m, x, y }$ such that $\vec\lambda \cdot \tup{ m, x, y } \in S$ for any method~$m$, argument-value~$x$, and admitted sequence $\vec\lambda \in S$. In particular, $\denot{W}_\mathrm{s}$ simulates any witnessing implementation $I$ whose abstract executions $E(I)$ are included in $E(\denot{W}_\mathrm{s})$\footnote{This is a consequence of a generic result stating that the set of traces of an LTS $A_1$ is included in the set of traces of an LTS $A_2$ iff $A_2$ simulates $A_1$, provided that $A_2$ is deterministic~\cite{DBLP:journals/iandc/LynchV95}.}. This completeness, however, extends only to inclusion of abstract executions, and not all the way to consistency, since consistency is defined on histories, and any given operation’s return value is not completely determined by the other operation labels and happens-before relation of a given history: return values generally depend on linearization order and visibility as well. Nevertheless, sequential specifications typically are return-value deterministic, and we have used simulation to prove consistency of Java-inspired weakly-consistent objects.

Establishing simulation for an implementation is also helpful when reasoning about clients of a concurrent object.
One can use the specification in place of the implementation and encode the client invariants using the abstract execution of the specification in order to prove client properties, following Sergey et al.’s approach~\cite{DBLP:conf/oopsla/SergeyNBD16}.

\subsection{Reducing Consistency to Safety Verification}
\label{sec:simulation:safety}

Proving simulation between an implementation and its specification can generally be achieved via product construction: complete the transition system of the specification, replacing non-enabled transitions with error-state transitions; then ensure the synchronized product of implementation and completed-specification transition systems is \emph{safe}, i.e.,~no error state is reachable. Assuming that the individual transition systems are safe, then the product system is safe \emph{iff} the specification simulates the implementation. This reduction to safety verification is also generally applicable to implementation and specification programs, though we limit our formalization to their underlying transition systems for simplicity. By Corollary~\ref{cor:sim}, such reductions enable consistency verification with existing safety verification tools.

\subsection{Verifying Implementation Programs}
\label{sec:simulation:programs}

While Theorem~\ref{thm:simulation} establishes forward simulation as a strategy for proving the consistency of implementations against weak-visibility specifications, its application to real-world implementations requires program-level mechanisms to signal the underlying AETS lin and vis actions.
To apply forward simulation, we thus develop a notion of programs whose commands include such mechanisms.

This section illustrates a toy programming language with AETS semantics which provides these mechanisms. The key features are the $\mathsf{lin}$ and $\mathsf{vis}$ program commands, which emit linearization and visibility actions for the currently-executing operation, along with $\mathsf{load}$, $\mathsf{store}$, and $\mathsf{cas}$ (compare-and-swap) commands, which record and return the set of operation identifiers having written to each memory cell. Such augmented memory commands allow programs to obtain handles to the operations whose effects it has observed, in order to signal the corresponding vis actions.

While one can develop similar mechanisms for languages with any underlying memory model, the toy language presented here assumes a sequentially-consistent memory.
Note that the assumption of sequentially-consistent memory operations is practically without loss of generality for Java’s concurrent collections since they are designed to be data-race free — their weak consistencies arise not from weak-memory semantics, but from non-atomic operations spanning several memory cells.

For generality, we assume abstract notions of commands and memory, using $\kappa$, $\mu$, $\ell$, and $M$ respectively to denote a \emph{program command}, \emph{memory command}, \emph{local state}, and \emph{global memory}. So that operations can assert their visibilities, we consider memory which stores, and returns upon access, the identifier(s) of operations which previously accessed a given cell. A \emph{program} $P = \tup{ \mathsf{init}, \mathsf{cmd}, \mathsf{idle}, \mathsf{done}}$ consists of an $\mathsf{init}(m, x) = \ell$ function mapping method name $m$ and argument values $x$ to local state $\ell$, along with a $\mathsf{cmd}(\ell) = \kappa$ function mapping local state $\ell$ to program command $\kappa$, and $\mathsf{idle}(\ell)$ and $\mathsf{done}(\ell)$ predicates on local states $\ell$. Intuitively, identifying local states with threads, the idle predicate indicates whether a thread is outside of atomic sections, and subject to interference from other threads; meanwhile the done predicate indicates whether whether a thread has terminated.

The \emph{denotation} of a memory command $\mu$ is a function $\denot{\mu}_\mathrm{m}$ from global memory $M_1$, argument value $x$, and operation $o$ to a tuple $\denot{\mu}_\mathrm{m}(M_1, x, o) = \tup{M_2, y}$ consisting of a global memory $M_2$, along with a return value $y$.

\begin{example}
  \label{ex:memory}

  A sequentially-consistent memory system which records the set of operations to access each location can be captured by mapping addresses $x$ to value and operation-set pairs $M(x) = \tup{y, O}$, along with three memory commands:
  \begin{align*}
    & \denot{\textsf{load}}_\mathrm{m}(M, x, \_) = \tup{M, M(x)}
    \qquad
    \denot{\textsf{store}}_\mathrm{m}(M, xy, o) = \tup{M[x \mapsto \tup{y,M(x)_1 \cup \set{o}}], \varepsilon}
    \\
    & \denot{\textsf{cas}}_\mathrm{m}(M, xyz, o) = \left\{
    \begin{array}{ll}
      \tup{M[x \mapsto \tup{z,M(x)_1 \cup \set{o}}], \mathit{true}} & \text{ if } M(x)_0 = y \\
      \tup{M, \mathit{false}} & \text{ if } M(x)_0 \neq y
    \end{array}
    \right.
  \end{align*}
  where the compare-and-swap (CAS) operation stores value $z$ at address $x$ and returns $\mathit{true}$ when $y$ was previously stored, and otherwise returns $\mathit{false}$.

\end{example}

The \emph{denotation} of a program command $\kappa$ is a function $\denot{\kappa}_\mathrm{c}$ from local state $\ell_1$ to a tuple $\denot{\kappa}_\mathrm{c}(\ell_1) = \tup{\mu, x, f}$ consisting of a memory command $\mu$ and argument value $x$, and a update continuation $f$ mapping the memory command’s return value $y$ to a pair $f(y) = \tup{\ell_2, \alpha}$, where $\ell_2$ is an updated local state, and $\alpha$ maps an operation $o$ to an LTS action $\alpha(o)$. We assume the denotation $\denot{{\tt ret}\ x}_\mathrm{c}(\ell_1) = \tup{\textsf{nop}, \varepsilon, \lambda y.\tup{\ell_2, \lambda o. \mathrm{ret}(z)}}$ of the {\tt ret} command yields a local state $\ell_2$ with $\mathsf{done}(\ell_2)$ without executing memory commands, and outputs a corresponding LTS ret action.

\begin{example}
  \label{ex:language}

  A simple goto language over variables ${\tt a}, {\tt b}, \ldots$ for the memory system of Example~\ref{ex:memory} would include the following commands:
  \begin{align*}
    \denot{\tt goto\ a}_\mathrm{c}(\ell)
    &=
    \tup{\textsf{nop}, \varepsilon, \lambda y. \tup{\mathit{jump}(\ell,\ell(a)), \lambda o.\varepsilon}}
    \\
    \denot{\tt assume\ a}_\mathrm{c}(\ell)
    &=
    \tup{\textsf{nop}, \varepsilon, \lambda y. \tup{\mathit{next}(\ell), \lambda o.\varepsilon}} \text{ if } \ell({\tt a}) \neq 0
    \\
    \denot{\tt b, c = load(a)}_\mathrm{c}(\ell)
    &=
    \tup{\textsf{load}, \ell({\tt a}), \lambda y_1, y_2. \tup{\mathit{next}(\ell[{\tt b} \mapsto y_1][{\tt c} \mapsto y_2]), \lambda o.\varepsilon}}
    \\
    \denot{\tt store(a,b)}_\mathrm{c}(\ell)
    &=
    \tup{\textsf{store}, \ell({\tt a})\ell({\tt b}), \lambda y. \tup{\mathit{next}(\ell), \lambda o.\varepsilon}}
    \\
    \denot{\tt d = cas(a,b,c)}_\mathrm{c}(\ell)
    &=
    \tup{\textsf{cas}, \ell({\tt a})\ell({\tt b})\ell({\tt c}), \lambda y. \tup{\mathit{next}(\ell[{\tt d} \mapsto y]), \lambda o.\varepsilon}}
  \end{align*}
  where the $\mathit{jump}$ and $\mathit{next}$ functions update a program counter, and the load command stores the operation identifier returned from the corresponding memory commands. Linearization and visibility actions are captured as program commands as follows:
  \begin{align*}
    \denot{\tt lin}_\mathrm{c}(\ell) &=  \tup{\textsf{nop}, \varepsilon, \lambda y. \tup{\mathit{next}(\ell), \lambda o.\mathrm{lin}(o)}} \\
    \denot{\tt vis(a)}_\mathrm{c}(\ell) &=  \tup{\textsf{nop}, \varepsilon, \lambda y. \tup{\mathit{next}(\ell), \lambda o.\mathrm{vis}(o,\ell({\tt a}))}}
  \end{align*}
  Atomic sections can be captured with a {\tt lock} variable and a pair of program commands,
  \begin{align*}
    \denot{\tt begin}_\mathrm{c}(\ell) &= \tup{\textsf{nop}, \varepsilon, \lambda y. \tup{\mathit{next}(\ell[{\tt lock} \mapsto \mathit{true}]), \lambda o.\varepsilon}} \\
    \denot{\tt end}_\mathrm{c}(\ell) &= \tup{\textsf{nop}, \varepsilon, \lambda y. \tup{\mathit{next}(\ell[{\tt lock} \mapsto \mathit{false}]), \lambda o.\varepsilon}}
  \end{align*}
  such that idle states are identified by not holding the lock, i.e.,~$\mathsf{idle}(\ell) = \lnot \ell({\tt lock})$, as in the initial state $\mathsf{init}(m, x)({\tt lock}) = \mathit{false}$.

\end{example}

Figure~\ref{fig:semantics} lists the semantics $\denot{P}_\mathrm{p}$ of a program $P$ as an abstract-execution transition system. The states $\tup{M, L}$ of $\denot{P}_\mathrm{p}$ include a global memory $M$, along with a partial function $L$ from operation identifiers $o$ to local states $L(o)$; the initial state is $\tup{ M_{\emptyset}, \emptyset }$, where $M_{\emptyset}$ is an initial memory state. The transitions for call and hb actions are enabled independently of implementation state, since they are dictated by implementations’ environments. Although we do not explicitly model client programs and platforms here, in reality, client programs dictate call actions, and platforms, driven by client programs, dictate hb actions; for example, a client which acquires the lock released after operation $o_1$, before invoking operation $o_2$, is generally ensured by its platform that $o_1$ happens before $o_2$. The transitions for all other actions are dictated by implementation commands. While the ret, lin, and vis commands generate their corresponding LTS actions, all other commands generate $\varepsilon$ transitions.

\begin{figure}
  \centering
  \begin{mathpar}
    \inferrule{
      o \not\in \dom(L) \\
      \ell = \mathsf{init}(m, x)
    }{
      \tup{M, L} \xrightarrow{\text{call}(o,m,x)} \tup{M, L[o \mapsto \ell]}
    }

    \inferrule{
      \mathsf{done}(L(o_1)) \\
      o_2 \not\in \dom(L)
    }{
      \tup{M, L} \xrightarrow{\text{hb}(o_1,o_2)} \tup{M, L}
    }

    \inferrule{
      \tup{M_1, \ell_1, o, \varepsilon} \leadsto^\ast \tup{M_2, \ell_2, o, \vec{a}} \\
      \mathsf{idle}(\ell_2)
    }{
      \tup{M_1, L[o \mapsto \ell_1]}
      \xrightarrow{\vec{a}}
      \tup{M_2, L[o \mapsto \ell_2]}
    }

    \inferrule{
      \mathsf{cmd}(\ell_1) = \kappa \\
      \denot{\kappa}_\mathrm{c}(\ell_1) = \tup{\mu, x, f} \\\\
      \denot{\mu}_\mathrm{m}(M_1, x, o) = \tup{M_2, y} \\
      f(y) = \tup{\ell_2, \alpha}
    }{
      \tup{M_1, \ell_1, o, \vec{a}}
      \leadsto
      \tup{M_2, \ell_2, o, \vec{a} \cdot \alpha(o)}
    }

  \end{mathpar}
  \caption{The semantics of program $P = \tup{ \mathsf{init}, \mathsf{cmd}, \mathsf{idle}, \mathsf{done} }$ as an abstract-execution transition system, where $\denot{\cdot}_\mathrm{c}$ and $\denot{\cdot}_\mathrm{m}$ are the denotations of program and memory commands, respectively.}
  \label{fig:semantics}

\end{figure}

Each atomic $\vecaioverarrow$ step of the AETS underlying a given program is built from a sequence of $\leadsto$ steps for the individual program commands in an atomic section. Individual program commands essentially execute one small $\leadsto$ step from shared memory and local state $\tup{M_1, \ell_1}$ to $\tup{M_2, \ell_2}$, invoking memory command $\mu$ with argument $x$, and emitting action $\alpha(o)$. Besides its effect on shared memory, each step uses the result $\tup{M_2, y}$ of memory command $\mu$ to update local state and emit an action using the continuation $f$, i.e.,~$f(y) = \tup{\ell_2, \alpha}$. Commands which do not access memory are modeled by a no-op memory commands. We define the consistency of programs by reduction to their transition systems.

\begin{definition}
  \label{def:consistent:program}

  A program $P$ is \emph{consistent} with a specification if{f} its semantics $\denot{P}_\mathrm{p}$ is.

\end{definition}

Thus the consistency of $P$ with $W$ amounts to the inclusion of $\denot{P}_\mathrm{p}$’s histories in $W$’s. The following corollary of Theorem~\ref{thm:simulation} follows directly by Definition~\ref{def:consistent:program}, and immediately yields a program verification strategy: validate a simulation relation from the states of $\denot{P}_\mathrm{p}$ to the states of $\denot{W}_\mathrm{s}$ such that each command of $P$ is simulated by a step of $\denot{W}_\mathrm{s}$.

\begin{corollary}
  \label{cor:sim}

  A program $P$ is consistent with specification $W$ if $\denot{W}_\mathrm{s}$ simulates $\denot{P}_\mathrm{p}$.

\end{corollary}


\section{Proof Methodology}
\label{sec:proving-refinement}

\begin{figure}[t]
	\centering
	\lstset{language=base, basicstyle=\scriptsize\ttfamily}
	\begin{minipage}{0.48\linewidth}
\begin{lstlisting}
var table: array of T;

procedure ^absolute^ put(k: int, v: T) {
  @atomic {@
    store(table[k],v);
    @vis(getLin());@
    |lin();|
  @}@
}
procedure ^absolute^ get(k: int) {
  @atomic{@
    v,O = load(table[k]);
    @vis(getLin());@
    |lin();|
  @}@
  return v;
}
\end{lstlisting}
	\end{minipage}
	\hfill
	\begin{minipage}{0.44\linewidth}
\begin{lstlisting}
procedure ^monotonic^ has(v: T)
  @vis(getModLin());@
{
  store(k,0);
  while ( k < table.length ) {
    @atomic{@
      tv, O = load(table[k]);
      @vis(O $\color{blue}{\cap}$ getModLin());@
    @}@
    if ( tv = v ) then
      |lin();|
      return true;
    inc(k);
  }
  |lin();|
  return false;
}
\end{lstlisting}
	\end{minipage}
	\vspace{-1em}
	\caption{An implementation~$I_\mathrm{chm}$ modeling Java’s concurrent hash map. The command \texttt{inc(k)} increments counter \texttt{k}, and commands within \texttt{atomic $\{\ldots\}$} are collectively atomic.}
	\label{fig:kvmap}
\end{figure}

%


In this section we develop a systematic means to annotating concurrent objects for relaxed-visibility simulation proofs. Besides leveraging an auxiliary memory system which tags memory accesses with the operations identifiers which wrote read values (see §\ref{sec:simulation:programs}), annotations signal linearization points with $\mathsf{lin}$ commands, and indicate visibility of other operations with $\mathsf{vis}$ commands. As in previous works~\cite{DBLP:conf/cav/AmitRRSY07, DBLP:conf/vmcai/Vafeiadis09, DBLP:journals/sttt/AbdullaHHJR17,DBLP:journals/toplas/HerlihyW90} we assume linearization points are given, and focus on visibility-related annotations.

As we focus on data-race free implementations (e.g., Java’s concurrent collections) for which sequential consistency is sound, it can be assumed without loss of generality that the happens-before order is exactly the \emph{returns-before} order between operations, which orders two operations $o_1$ and $o_2$ iff the return action of $o_1$ occurs in real-time before the call action of $o_2$. This assumption allows to guarantee that linearizations are consistent with happens-before just by ensuring that the linearization point of each operation occurs in between its call and return action (like in standard linearizability). It is without loss of generality because the clients of such implementations can use auxiliary variables to impose synchronization order constraints between every two operations ordered by returns-before, e.g., writing a variable after each operation returns which is read before each other operation is called (under sequential consistency, every write happens-before every other read which reads the written value). 

We illustrate our methodology with the key-value map implementation~$I_\mathrm{chm}$ of Figure~\ref{fig:kvmap}, which models Java’s concurrent hash map. The lines marked in blue and red represent linearization/visibility commands added by the instrumentation that will be described below. Key-value pairs are stored in an array \texttt{table} indexed by keys.  The implementation of \texttt{put} and \texttt{get} are obvious
while the implementation of \texttt{has} returning true iff the input value is associated to some key consists of a while loop traversing the array and searching for the input value. To simplify the exposition, the shared memory reads and writes are already adapted to the memory system described in Section~\ref{sec:simulation:programs} (essentially, this consists in adding new variables storing the set of operation identifiers returned by a shared memory read).
While \texttt{put} and \texttt{get} are obviously linearizable, \texttt{has} is weakly consistent, with monotonic visibility. For instance, given the two thread program
$
 {\tt \{get(1);has(1)\}\ ||\ \{put(1,1);put(0,1);put(1,0)\}}
$
it is possible that \texttt{get(1)} returns 1 while \texttt{has(1)} returns false. This is possible in an interleaving where \texttt{has} reads \texttt{table}[0] before \texttt{put(0,1)}  writes into it (observing the initial value 0), and \texttt{table}[1] after \texttt{put(1,0)} writes into it (observing value 0 as well). The only abstract execution consistent with the weakly-consistent contains-value map $W_{\mathrm{m}}$ (Example~\ref{ex:spec:weak}) which justifies these return values is given in Example~\ref{ex:execution}.
We show that this implementation is consistent with a simplification of the contains-value map $W_{\mathrm{m}}$, without remove key operations, and where put operations return no value.

Given an implementation $I$, let $\mathcal{L}(I)$ be an instrumentation of $I$ with program commands \textcolor{red}{\tt lin()} emitting linearization actions. The execution of \textcolor{red}{\tt lin()} in the context of an operation with identifier $o$ emits a linearization action $\textrm{lin}(o)$. We assume that $\mathcal{L}(I)$ leads to well-formed executions (e.g., at most one linearization action per operation).

\begin{example}

For the implementation in Figure~\ref{fig:kvmap}, the linearization commands of \texttt{put} and \texttt{get} are executed atomically with the store to \texttt{table[k]} in \texttt{put} and the load of \texttt{table[k]} in \texttt{get}, respectively. The linearization command of \texttt{has} is executed at any point after observing the input value \texttt{v} or after exiting the loop, but before the return. The two choices correspond to different return values and only one of them will be executed during an invocation.

\end{example}


Given an instrumentation $\mathcal{L}(I)$, a visibility annotation $V$ for $I$'s methods, and a read-only predicate $R$, we define a witnessing implementation $\mathcal{V}(\mathcal{L}(I))$ according to a generic heuristic that depends only on $V$ and $R$. This definition uses a program command \texttt{getLin()} which returns the set of operations in the current linearization sequence\footnote{We rely on retrieving the identifiers of currently-linearized operations. More complex proofs may also require inspecting, e.g.,~operation labels and happens-before relationships.}. The current linearization sequence is stored in a history variable which is updated with every linearization action by appending the corresponding operation identifier. For readability, we leave this history variable implicit and omit the corresponding updates. As syntactic sugar, we use a command \texttt{getModLin()} which returns the set of \emph{modifiers} (non read-only operations) in the current linearization sequence. To represent visibility actions, we use program commands \textcolor{blue}{\texttt{vis(A)}} where $\texttt{A}$ is a set of operation identifiers. The execution of \textcolor{blue}{\texttt{vis(A)}} in the context of an operation with identifier $o$ emits the set of visibility actions $\mathrm{vis}(o,o')$ for every operation $o'\in \texttt{A}$.

Therefore, $\mathcal{V}(\mathcal{L}(I))$ extends the instrumentation $\mathcal{L}(I)$ with commands generating visibility actions as follows:
\begin{itemize}
\item for absolute methods, each linearization command is preceded by \textcolor{blue}{\texttt{vis(getLin())}} which ensures that the visibility of an invocation includes all the predecessors in linearization order. This is executed atomically with \textcolor{red}{\tt lin()}.
\item for monotonic methods, the call action is followed by \textcolor{blue}{\texttt{vis(getModLin())}} (and executed atomically with this command) which ensures that the visibility of each invocation is monotonic, and every read of a shared variable which has been written by a set of operations $O$ is preceded by \textcolor{blue}{\texttt{vis(O $\cap$ getModLin())}} (and executed atomically with this  command). The latter is needed so that the visibility of such an invocation contains enough operations to explain its return value (the visibility command attached to call actions is enough to ensure monotonic visibilities).
\end{itemize}


\begin{example}

The blue lines in Figure~\ref{fig:kvmap} demonstrate the visibility commands added by the instrumentation $\mathcal{V}(\cdot)$ to the key-value map in Figure~\ref{fig:kvmap} (in this case, the modifiers are \texttt{put} operations). The first visibility command in \texttt{has} precedes the procedure body to emphasize the fact that it is executed \emph{atomically} with the procedure call. Also, note that the read of the array \texttt{table} is the only shared memory read in \texttt{has}.

\end{example}

\begin{theorem}

The abstract executions of the witnessing implementation $\mathcal{V}(\mathcal{L}(I))$ are consistent with $V$ and $R$.

\end{theorem}
\begin{proof}

Let $\tup{h, \mathit{lin}, \mathit{vis}}$ be the abstract execution of a trace $\tau$ of $\mathcal{V}(\mathcal{L}(I))$, and let $o$ be an invocation in $h$ of a monotonic method (w.r.t. $V$). By the definition of $\mathcal{V}$, the call action of $o$ is \emph{immediately} followed in $\tau$ by a sequence of visibility actions $\mathrm{vis}(o,o')$ for every modifier $o'$ which has been already linearized.
Therefore, any operation which has returned before $o$ (i.e., happens-before $o$) has already been linearized and it will necessarily have a smaller visibility (w.r.t. set inclusion) because the linearization sequence is modified only by appending new operations. The instrumentation of shared memory reads may add more visibility actions $\mathrm{vis}(o,\_)$ but this preserves the monotonicity status of $o$'s visibility. The case of absolute methods is obvious.
\end{proof}

The consistency of the abstract executions of $\mathcal{V}(\mathcal{L}(I))$ with a given sequential specification $S$, which completes the proof of consistency with a weak-visibility specification $W=\tup{S,R,V}$,
can be proved by showing that the transition system $\denot{W}_\mathrm{s}$ of $W$ simulates $\mathcal{V}(\mathcal{L}(I))$ (Theorem~\ref{thm:simulation}).
Defining a simulation relation between the two systems is in some part implementation specific, and in the following we demonstrate it for the key-value map implementation $\mathcal{V}(\mathcal{L}(I_\mathrm{chm}))$.

We show that $\denot{W_\mathrm{m}}_\mathrm{s}$ simulates implementation~$I_\mathrm{chm}$. A state of $I_\mathrm{chm}$ in Figure~\ref{fig:kvmap} is a valuation of \texttt{table} and the history variable \texttt{lin} storing the current linearization sequence, and a valuation of the local variables for each active operation. Let $\mathit{ops}(q)$ denote the set of operations which are active in an implementation state $q$. Also, for a \texttt{has} operation $o\in \mathit{ops}(q)$, let $\mathit{index}(o)$ be the maximal index $k$ of the array \texttt{table} such that $o$ has already read $\texttt{table}[k]$ and $\texttt{table}[k]\neq \texttt{v}$. We assume $\mathit{index}(o)=-1$ if $o$ did not read any array cell.

\begin{definition}
  \label{def:array-map-simulation-relation}
Let $R_\mathrm{chm}$ be a relation which associates every implementation state $q$ with a state of $\denot{W_\mathrm{m}}_\mathrm{s}$, i.e., an $\tup{S,R,V}$-consistent abstract execution $e=\tup{h, \mathit{lin}, \mathit{vis}}$ with $h = \tup{O, \mathit{inv}, \mathit{ret}, \mathit{hb}}$, such that:
\begin{enumerate}
	\item $O$ is the set of identifiers occurring in $\mathit{ops}(q)$ or the history variable \emph{\texttt{lin}},
	\item for each operation $o\in \mathit{ops}(q)$, $\mathit{inv}(o)$ is defined according to its local state, $\mathit{ret}(o)$ is undefined, and $o$ is maximal in the happens-before order $\mathit{hb}$,
	\item\label{it:unspec1} the value of the history variable \emph{\texttt{lin}} in $q$ equals the linearization sequence $\mathit{lin}$,
	\item\label{it:unspec2} every invocation $o\in \mathit{ops}(q)$ of an absolute method (\texttt{\emph{put}} or \texttt{\emph{get}}) has absolute visibility if linearized, otherwise, its visibility is empty,
	\item\label{it:spec1} \texttt{\emph{table}} is the array obtained by executing the sequence of operations $\mathit{lin}$,
	\item\label{it:spec2} for every linearized \texttt{\emph{get(k)}} operation $o\in \mathit{ops}(q)$, the \texttt{\emph{put(k,\_)}} operation in $\mathit{vis}(o)$ which occurs last in  $\mathit{lin}$ writes \texttt{\emph{v}} to key \texttt{\emph{k}}, where \texttt{\emph{v}} is the local variable of $o$,
	\item\label{it:spec3} for every \texttt{\emph{has}} operation $o\in \mathit{ops}(q)$, 
	$\mathit{vis}(o)$ consists of:
	\begin{itemize}
		\item all the \texttt{\emph{put}} operations $o'$ which returned before $o$ was invoked,
		\item for each $i\leq \mathit{index}(o)$, all the \texttt{\emph{put($i$,\_)}} operations from a prefix of $\mathit{lin}$ that wrote a value \emph{different} from $\texttt{v}$,
		\item all the \texttt{\emph{put($\mathit{index}(o)+1$,\_)}} operations from a prefix of $\mathit{lin}$ that ends with a \texttt{\emph{put($\mathit{index}(o)+1$,v)}} operation, provided that $\texttt{\emph{tv}}=\texttt{\emph{v}}$.
	\end{itemize}
	Above, the linearization prefix associated to an index $j_1<j_2$ should be a prefix of the one associated to $j_2$.
\end{enumerate}
\end{definition}

A large part of this definition is applicable to any implementation, only points (\ref{it:spec1}), (\ref{it:spec2}), and (\ref{it:spec3}) being specific to the implementation we consider. The points (\ref{it:spec2}) and (\ref{it:spec3}) ensure that the return values of operations are consistent with $S$ and mimic the effect of the \texttt{\emph{vis}} commands from Figure~\ref{fig:kvmap}.

\begin{theorem}
  \label{thm:simulation:map}

  $R_\mathrm{chm}$ is a simulation relation from $\mathcal{V}(\mathcal{L}(I_\mathrm{chm}))$ to $\denot{W_{\mathrm{m}}}_\mathrm{s}$.

\end{theorem}


\newcommand{\callAction}{\textnormal{\textrm{call}}}
\newcommand{\retAction}{\textnormal{\textrm{ret}}}
\newcommand{\hbAction}{\textnormal{\textrm{hb}}}
\newcommand{\visAction}{\textnormal{\textrm{vis}}}
\newcommand{\linAction}{\textnormal{\textrm{lin}}}

\newcommand{\atomicSequences}{\mathsf{AtomicSeq}}

\section{Implementation and Evaluation}
\label{sec:implementation}


In this section we effectuate our methodology by verifying two weakly-consistent concurrent objects: Java’s \code{ConcurrentHashMap} and \code{ConcurrentLinkedQueue}.\footnote{Our verified implementations are open source, and available at:\\ \url{https://github.com/siddharth-krishna/weak-consistency-proofs}.}
We use an off-the-shelf deductive verification tool called \civl{}~\cite{DBLP:conf/cav/HawblitzelPQT15}, though any concurrent program verifier could suffice.
We chose \civl{} because comparable verifiers either require a manual encoding of the concurrency reasoning (e.g. Dafny or Viper) which can be error-prone, or require cumbersome reasoning about interleavings of thread-local histories (e.g. VerCors).
An additional benefit of \civl{} is that it directly proves simulation, thereby tying the mechanized proofs to our theoretical development.
Our proofs assume no bound on the number of threads or the size of the memory.

Our use of \civl{} imposes two restrictions on the implementations we can verify.
First, \civl{} uses the Owicki-Gries method~\cite{DBLP:journals/cacm/OwickiG76} 
to verify concurrent programs.
These methods are unsound for weak memory models~\cite{DBLP:conf/icalp/LahavV15}, so \civl{}, and hence our proofs, assume a sequentially-consistent memory model.
Second, \civl{}'s strategy for building the simulation relation requires implementations to have statically-known linearization points because it checks that there exists exactly one atomic section in each code path where the global state is modified, and this modification is simulated by the specification.

Given these restrictions, we can simplify our proof strategy of forward refinement by factoring the simulations we construct through an atomic version of the specification transition system.
This atomic specification is obtained from the specification AETS $\denot{W}_\mathrm{s}$ by  restricting the interleavings between its transitions. 


\begin{definition}

  The \emph{atomic transition system} of a specification $W$ is the AETS $\denot{W}_\mathrm{a} = \tup{Q, A, q, \to_a}$, where $\denot{W}_\mathrm{s} = \tup{Q, A, q, \to}$ is the AETS of $W$ and $e_1 \xrightarrow{\vec{a}}_a e_2$ if and only if $e_1 \xrightarrow{\vec{a}} e_2$ and $\vec{a} \in \set{\callAction(o, m, x)} \cup \set{\retAction(o, y)} \cup \set{\hbAction(o, o')} \cup \set{\vec{a_1} \; \linAction(o) : \vec{a_1} \in \set{\visAction(o, \_)}^*}$.

\end{definition}
Note that the language of $\denot{W}_\mathrm{a}$ is included in the language of $\denot{W}_\mathrm{s}$ and simulation proofs towards $\denot{W}_\mathrm{a}$ apply to $\denot{W}_\mathrm{s}$ as well.



Our \civl{} proofs show that there is a simulation from an implementation to its atomic specification, which
is encoded as a program whose state consists of the components of an abstract execution, i.e., $\tup{O, \mathit{inv}, \mathit{ret}, \mathit{hb}, \mathit{lin}, \mathit{vis}}$.
These were encoded as maps from operation identifiers to values, sequences of operation identifiers, and maps from operation identifiers to sets of operation identifiers respectively.
Our axiomatization of sequences and sets were adapted from those used by the Dafny verifier~\cite{DBLP:conf/lpar/Leino10}.
For each method in $\mathbb{M}$, we defined atomic procedures corresponding to call actions, return actions, and combined visibility and linearization actions in order to obtain exactly the atomic transitions of $\denot{W}_\mathrm{a}$.

It is challenging to encode Java implementations faithfully in \civl{}, as the latter's input programming language is a basic imperative language lacking many Java features.
Most notable among these is dynamic memory allocation on the heap, used by almost all of the concurrent data structure implementations.
As \civl{} is a first-order prover, we needed an encoding of the heap that lets us perform reachability reasoning on the heap.
We adapted the first-order theory of reachability and footprint sets from the GRASShopper verifier~\cite{DBLP:conf/tacas/PiskacWZ14} 
for dynamically allocated data structures.
This fragment is decidable, but relies on local theory extensions~\cite{DBLP:conf/cade/Sofronie-Stokkermans05}, 
which we implemented by using the trigger mechanism of the underlying SMT solver \cite{DBLP:journals/entcs/MoskalLK08, DBLP:journals/corr/abs-1004-3808} to ensure that quantified axioms were only instantiated for program expressions.
For instance, here is the ``cycle'' axiom that says that if a node \code{x} has a field \code{f[x]} that points to itself, then any \code{y} that it can reach via that field (encoded using the between predicate \code{Btwn(f, x, y, y)}) must be equal to \code{x}:
\begin{lstlisting}[language=boogie]
  axiom (forall f: [Ref]Ref, x: Ref, y:Ref :: {known(x), known(y)}
    f[x] == x && Btwn(f, x, y, y) ==> x == y); 
\end{lstlisting}
We use the trigger \code{{known(x), known(y)}} (\code{known} is a dummy function that maps every reference to true) and introduce \code{known(t)} terms in our programs for every term \code{t} of type \code{Ref} (for instance, by adding \code{assert known(t)} to the point of the program where \code{t} is introduced).
This ensures that the cycle axiom is only instantiated for terms that appear in the program, and not for terms that are generated by instantations of axioms (like \code{f[x]} in the cycle axiom).
This process was key to keeping the verification time manageable.

Since we consider fine-grained concurrent implementations, we also needed to reason about interference by other threads and show thread safety.
\civl{} provides Owicki-Gries~\cite{DBLP:journals/cacm/OwickiG76} style thread-modular reasoning, by means of demarcating atomic blocks and providing preconditions for each block that are checked for stability under all possible modifications by other threads.
One of the consequences of this is that these annotations can only talk about the local state of a thread and the shared global state, but not other threads.
To encode facts such as distinctness of operation identifiers and ownership of unreachable nodes (e.g. newly allocated nodes) in the shared heap, we use \civl{}'s linear type system ~\cite{DBLP:conf/ifip2/Wadler90}.

For instance, the proof of the \code{push} method needs to make assertions about the value of the newly-allocated node \code{x}.
These assertions would not be stable under interference of other threads if we didn't have a way of specifying that the address of the new node is known only by the \code{push} thread.
We encode this knowledge by marking the type of the variable \code{x} as linear -- this tells \civl{} that all values of \code{x} across all threads are distinct, which is sufficient for the proof.
\civl{} ensures soundness by making sure that linear variables are not duplicated (for instance, they cannot be passed to another method and then used afterwards).

\begin{figure}[t]
  \centering
  \def\arraystretch{0.8}
  \begin{tabular}{l r r r r}
    \toprule
    \bf Module & \bf Code & \bf Proof & \bf Total & \bf Time (s) \\
    \midrule
    Sets and Sequences & - & 85 & 85 & - \\
    Executions and Consistency & - & 30 & 30 & - \\
    Heap and Reachability & - & 35 & 35 & - \\
    \midrule
    Map ADT & 51 & 34 & 85 & - \\
    Array-map implementation & 138 & 175 & 313 & 6 \\
    \midrule
    Queue ADT & 50 & 22 & 72 & - \\
    Linked Queue implementation & 280 & 325 & 605 & 13 \\
    \bottomrule
  \end{tabular}
  \caption{Case study detail: for each object we show lines of code, lines of proof, total lines, and verification time in seconds. We also list common definitions and axiomatizations separately.}
  \label{tab:case-studies}
\end{figure}

We evaluate our proof methodology by considering models of two of Java's weakly-consistent concurrent objects.

\subsubsection{Concurrent Hash Map}

One is the \code{ConcurrentHashMap} implementation of the Map ADT, consisting of absolute \code{put} and \code{get} methods and a monotonic \code{has} method that follows the algortihm given in Figure~\ref{fig:kvmap}.
For simplicity, we assume keys are integers and the hash function is identity, but note that the proof of monotonicity of \code{has} is not affected by these assumptions.

\civl{} can construct a simulation relation equivalent to the one defined in Definition~\ref{def:array-map-simulation-relation} automatically, given an inductive invariant that relates the state of the implementation to the abstract execution.
A first attempt at an invariant might be that the value stored at \code{table[k]} for every key \code{k} is the same as the value returned by adding a \code{get} operation on \code{k} by the specification AETS.
This invariant is sufficient for \civl{} to prove that the return value of the absolute methods (\code{put} and \code{get}) is consistent with the specification.

However, it is not enough to show that the return value of the monotonic \code{has} method is consistent with its visibility.
This is because our proof technique constructs a visibility set for \code{has} by taking the union of the memory tags (the set of operations that wrote to each memory location) of each table entry it reads,  but without additional invariants this visibility set could entail a different return value.
We thus strengthen the invariant to say that \code{tableTags[k]}, the memory tags associated with hash table entry \code{k}, is exactly the set of linearized \code{put} operations with key \code{k}.
A consequence of this is that the abstract state encoded by \code{tableTags[k]} has the same value for key \code{k} as the value stored at \code{table[k]}.
\civl{} can then prove, given the following loop invariant, that the value returned by \code{has} is consistent with its visibility set.
\begin{lstlisting}[language=boogie]
(forall i: int :: 0 <= i && i < k ==> Map.ofVis(my_vis, lin)[i] != v)
\end{lstlisting}
This loop invariant says that among the entries scanned thus far, the abstract map given by the projection of \code{lin} to the current operation's visibility \code{my\_vis} does not include value \code{v}.

\subsubsection{Concurrent Linked Queue}

\begin{figure}[t]
  \centering
  \lstset{language=base, basicstyle=\scriptsize\ttfamily}
  \begin{minipage}[t]{0.48\linewidth}
\begin{lstlisting}
var head: Ref;
var tail: Ref;

struct Node {
  var data: K;
  var next: Ref;
}

procedure ^absolute^ push(k: K) {
  x := new Node(k, null);
  while (true) {
    t, _ := load(tail);
    tn, _ := load(tail.next);
    if (tn == null) {
      @atomic{@
        b, _ := cas(t.next, tn, x);
        if (b) {
          @vis(getLin());@
          |lin();|
        }
      @}@
      if (b) then break;
    } else {
      b, _ := cas(tail, t, tn);
    }
  }
}
\end{lstlisting}
  \end{minipage}
  \begin{minipage}[t]{0.48\linewidth}
\begin{lstlisting}
procedure ^absolute^ pop() {
  while (true) {
    h, _ := load(head);
    t, _ := load(tail);
    hn, _ := load(h.next);
    if (h != t) {
      k, _ := load(hn.data);
      @atomic{@
        b, _ := cas(head, h, hn);
        if (b) {
          @vis(getLin());@
          |lin();|
        }
      @}@
      if (b) then return k;
}}}

proccedure ^monotonic^ size()
  @vis(getModLin());@
{
  s := 0;
  c, _ := load(head);
  @atomic{@
    cn, O := load(c.next);
    @vis(O $\color{blue}{\cap}$ getModLin());@
  @}@  
  while (cn != null) {
    s := s + 1;
    c := cn;
    @atomic{@
      cn, O := load(c.next);
      @vis(O $\color{blue}{\cap}$ getModLin());@
    @}@  
  }
  |lin();|
  return s;
}
\end{lstlisting}
  \end{minipage}
  \caption{The simplified implementation of Java's \code{ConcurrentLinkedQueue} that we verify.}
  \label{fig:queue}
\end{figure}

Our second case study is the \code{ConcurrentLinkedQueue} implementation of the Queue ADT, consisting of absolute \code{push} and \code{pop} methods and a monotonic \code{size} method that traverses the queue from head to tail without any locks and returns the number of nodes it sees (see Figure~\ref{fig:queue} for the full code).
We again model the core algorithm (the Michael-Scott queue~\cite{DBLP:conf/podc/MichaelS96}) and omit some of Java's optimizations, for instance to speed up garbage collection by setting the \code{next} field of popped nodes to themselves, or setting the values of nodes to \code{null} when popping values.

The invariants needed to verify the absolute methods are a straightforward combination of structural invariants (e.g. that the queue is composed of a linked list from the head to null, with the tail being a member of this list) and a relation between the abstract and concrete states.
Once again, we need to strengthen this invariant in order to verify the monotonic \code{size} method, because otherwise we cannot prove that the visibility set we construct (by taking the union of the memory tags of nodes in the list during traversal) justifies the return value.

The key additional invariant is that the memory tags for the next field of each node (denoted \code{x.nextTags} for each node \code{x}) in the queue contain the operation label of the operation that pushed the next node into the queue (if it exists).
Further, the sequence of operations in \code{lin} are exactly the operations in the \code{nextTags} field of nodes in the queue, and in the order they are present in the queue.
Given this invariant, one can show that the return value \code{s} computed by \code{size} is consistent with the visibility set it constructs by picking up the memory tags from each node that it traverses (the loop invariant is more involved, as due to concurrent updates \code{size} could be traversing nodes that have been popped from the queue).

\subsubsection{Results}

Figure~\ref{tab:case-studies} provides a summary of our case studies.
We separate the table into sections, one for each case study, and a common section at the top that contains the common theories of sets and sequences and our encoding of the heap.
In each case study section, we separate the definitions of the atomic specification of the ADT (which can be reused for other implementations) from the code and proof of the implementation we consider.
For each resulting module, we list the number of lines of code, lines of proof, total lines, and \civl{}'s verification time in seconds.
Experiments were conducted on an Intel Core i7-4470 3.4 GHz 8-core machine with 16GB RAM.

Our two case studies are representative of the weakly-consistent behaviors exhibited by all the Java concurrent objects studied in~\cite{DBLP:journals/pacmpl/EmmiE19}, both those using fixed-size arrays and those using dynamic memory.
As \civl{} does not direclty support dynamic memory and other Java language features, we were forced to make certain simplifications to the algorithms in our verification effort.
However, the assumptions we make are orthogonal to the reasoning and proof of weak consistency of the monotonic methods.
The underlying algorithm used by, and hence the proof argument for monotonicity of, hash map's \code{has} method is the same as that in the other monotonic hash map operations such as \code{elements}, \code{entrySet}, and \code{toString}.
Similarly, the argument used for the queue's \code{size} can be adapted to other monotonic \code{ConcurrentLinkedQueue} and \code{LinkedTransferQueue} operations like \code{toArray} and \code{toString}.
Thus, our proofs carry over to the full versions of the implementations as the key invariants linking the memory tags and visibility sets to the specification state are the same.

In addition, \civl{} does not currently have any support for inferring the preconditions of each atomic block, which currently accounts for most of the lines of proof in our case studies.
However, these problems have been studied and solved in other tools~\cite{DBLP:conf/tacas/PiskacWZ14, DBLP:conf/vmcai/Vafeiadis10}, and in theory can be integrated with \civl{} in order to simplify these kinds of proofs.

In conclusion, our case studies show that verifying weakly-consistent operations introduces little overhead compared to the proofs of the core absolute operations.
The additional invariants needed to prove monotonicity were natural and easy to construct.
We also see that our methodology brings weak-consistency proofs within the scope of what is provable by off-the-shelf automated concurrent program verifiers in reasonable time.










\section{Related Work}
\label{sec:related-work}

Though \emph{linearizability}~\cite{DBLP:journals/toplas/HerlihyW90} has reigned as the de-facto concurrent-object consistency criterion, several recent works proposed weaker criteria, including 
\emph{quantitative relaxation}~\cite{DBLP:conf/popl/HenzingerKPSS13}, \emph{quiescent consistency}~\cite{DBLP:conf/fm/DerrickDSTTW14}, 
and
\emph{local linearizability}~\cite{DBLP:conf/concur/HaasHHKLPSSV16}; 
these works effectively permit externally-visible interference among threads by altering objects’ sequential specifications, each in their own way. Motivated by the diversity of these proposals, Sergey et al.~\cite{DBLP:conf/oopsla/SergeyNBD16} proposed the use of Hoare logic 
for describing a custom consistency specification for each concurrent object. Raad et al.~\cite{DBLP:journals/pacmpl/RaadDRLV19} continued in this direction by proposing declarative consistency models for concurrent objects atop weak-memory platforms.
One common feature between our paper and this line of work (see also~\cite{DBLP:conf/esop/KhyzhaDGP17, DBLP:conf/ecoop/DelbiancoSNB17}) is encoding and reasoning directly about the concurrent history.
The notion of \emph{visibility relaxation}~\cite{DBLP:journals/pacmpl/EmmiE19} originates from Burckhardt et al.’s axiomatic specifications~\cite{DBLP:conf/popl/BurckhardtGYZ14}, and leverages traditional sequential specifications by allowing certain operations to behave as if they are unaware of concurrently-executed linearization-order predecessors. The linearization (and visibility) actions of our simulation-proof methodology are unique to visibility-relaxation based weak-consistency, since they refer to a global linearization order linking executions with sequential specifications.

Typical methodologies for proving linearizability are based on reductions to safety verification~\cite{DBLP:journals/corr/ChakrabortyHSV15, DBLP:journals/iandc/BouajjaniEEH18} and forward simulation~\cite{DBLP:conf/cav/AmitRRSY07, DBLP:conf/vmcai/Vafeiadis09, DBLP:journals/sttt/AbdullaHHJR17}, 
the latter generally requiring the annotation of per-operation \emph{linearization points}, each typically associated with a single program statement in the given operation, e.g.,~a shared memory access. Extensions to this methodology include \emph{cooperation}~\cite{DBLP:conf/cav/Vafeiadis10, DBLP:conf/cav/DragoiGH13, DBLP:conf/cav/ZhuPJ15}, 
i.e.,~allowing operations’ linearization points to coincide with other operations’ statements, and \emph{prophecy}~\cite{DBLP:conf/cav/SchellhornWD12, DBLP:conf/pldi/LiangF13}, 
i.e.,~allowing operation’ linearization points to depend on future events. Such extensions enable linearizability proofs of objects like the Herlihy-Wing Queue (HWQ). 
While prophecy~\cite{DBLP:journals/iandc/LynchV95}, 
alternatively backward simulation~\cite{DBLP:journals/iandc/LynchV95}, is generally more powerful than forward simulation alone, Bouajjani et al.~\cite{DBLP:conf/cav/BouajjaniEEM17} described a methodology based on forward simulation capable of proving seemingly future-dependent objects like HWQ by considering fixed linearization points only for value removal, and an additional kind of specification-simulated action, \emph{commit points}, corresponding to operations’ final shared-memory accesses. Our consideration of specification-simulated visibility actions follows this line of thinking, enabling the forward-simulation based proof of weakly-consistent concurrent objects.

  \section{Conclusion and Future Work}
\label{sec:conclusion}

This work develops the first verification methodology for weakly-consistent operations using sequential specifications and forward simulation, thus reusing existing sequential ADT specifications and enabling simple reasoning, i.e.,~without prophecy~\cite{DBLP:journals/tcs/AbadiL91} or backward simulation~\cite{DBLP:journals/iandc/LynchV95}.
While we have already demonstrated the application to absolute and monotonic methods on sequentially-consistent memory, our formalization is general, and also applicable to the other visibility relaxations, e.g.,~the \emph{peer} and \emph{weak} visibilities~\cite{DBLP:journals/pacmpl/EmmiE19}, and weaker memory models, e.g.,~the Java memory model.
These extensions amount to devising additional visibility-action inference strategies (§\ref{sec:proving-refinement}), and alternate memory-command denotations (§\ref{sec:simulation:programs}).

As with systematic or automated linearizability-proof methodologies, our proof methodology is susceptible to two potential sources of incompleteness. First, as mentioned in Section~\ref{sec:simulation}, methodologies like ours based on forward simulation are only complete when specifications are \emph{return-value deterministic}. 
However, data types are typically designed to be return-value deterministic and this source of incompleteness does not manifest in practice.

Second, methodologies like ours based on annotating program commands, e.g.,~with linearization points, are generally incomplete since the consistency mechanism employed by any given implementation may not admit characterization according to a given static annotation scheme; the Herlihy-Wing Queue, whose linearization points depend on the results of future actions, is a prototypical example~\cite{DBLP:journals/toplas/HerlihyW90}. Likewise, our systematic strategy for annotating implementations with \emph{lin} and \emph{vis} commands (§\ref{sec:simulation}) can fail to prove consistency of future-dependent operations. However, we have yet to observe any practical occurrence of such exotic objects; our strategy is sufficient for verifying the weakly-consistent algorithms implemented in the Java development kit. As a theoretical curiosity for future work, investigating the potential for complete annotation strategies would be interesting, e.g.,~for restricted classes of data types and/or implementations.

Finally, while \civl{}’s high-degree of automation facilitated rapid prototyping of our simulation proofs, its underlying foundation using Owicki-Gries style proof rules limits the potential for modular reasoning. In particular, while our weak-consistency proofs are thread-modular, our invariants and intermediate assertions necessarily talk about state shared among multiple threads. Since our simulation-based methodology and annotations are completely orthogonal to the underlying program logic, it would be interesting future work to apply our methodology using expressive logics like Rely-Guarantee, e.g.~\cite{DBLP:conf/ifip/Jones83,DBLP:conf/cav/Vafeiadis10}, or variations of Concurrent Separation Logic, e.g.~\cite{DBLP:conf/concur/OHearn04,DBLP:conf/lics/Reynolds02,DBLP:conf/pldi/SergeyNB15,DBLP:conf/oopsla/SergeyNBD16,DBLP:conf/ifm/BlomDHO17,DBLP:journals/jfp/JungKJBBD18}. It remains to be seen to what degree increased modularity may sacrifice automation in the application of our weak-consistency proof methodology.


  \bibliography{dblp,misc}

  \newpage
  \appendix

\section{Proofs to Theorems and Lemmas}
\label{sec:proofs}

\begin{replemma}{lem:executions}

  The abstract executions~$E(W)$ of a specification~$W$ are consistent with $W$.

\end{replemma}

\begin{proof}

  Any complete, sequential, and absolute execution is consistent by definition, since the labeling of its linearization is taken from the sequential specification. Then, any happens-before weakening is consistent for exactly the same reason as its source execution, since its linearization and visibility projection are both identical. Finally, any visibility weakening is consistent by the condition of $W$-consistency in its definition.
\end{proof}

\begin{replemma}{lem:histories}

  A weak-visibility specification and its transition system have identical histories.

\end{replemma}

\begin{proof}

  It follows almost immediately that the abstract executions of $\denot{W}_\mathrm{s}$ are identical to those of $W$, since $\denot{W}_\mathrm{s}$’s state effectively records the abstract execution of a given AETS execution, and only enables those returns that are consistent with $W$. Since histories are the projections of abstract executions, the corresponding history sets are also identical.
\end{proof}

\begin{reptheorem}{thm:simulation}

  A witnessing implementation~$I$ is consistent with a weak-visibility specification~$W$ if the transition system $\denot{W}_\mathrm{s}$ of $W$ simulates $I$.

\end{reptheorem}

\begin{proof}

  This follows from standard arguments, given that the corresponding SLTSs include $\varepsilon$ transitions to ensure that every move of one system can be matched by stuttering from the other: since both systems synchronize on the call, ret, hb, lin, and vis actions, the simulation guarantees that every abstract execution, and thus history, of $I$ is matched by one of $\denot{W}_\mathrm{s}$. Then by Lemma~\ref{lem:histories}, the histories of $I$ are included in $W$.
\end{proof}

\begin{reptheorem}{thm:simulation:map}

  $R_\mathrm{chm}$ is a simulation relation from $I_\mathrm{chm}$ to $\denot{W_{\mathrm{m}}}_\mathrm{s}$.

\end{reptheorem}

\begin{proof}[Sketch]

  We show that every step of the implementation, i.e., an atomic section or a program command, is simulated by $\denot{W_\mathrm{m}}_\mathrm{s}$. Given $\tup{q,e}\in R_\mathrm{chm}$, we consider the different implementation steps which are possible in $q$.

  The case of commands corresponding to procedure calls of \texttt{put} and \texttt{get} is trivial. Executing a procedure call in $q$ leads to a new state $q'$ which differs only by having a new active operation $o$. We have that $e \xrightarrow{\textrm{call}(o,\_,\_)} e'$ and $\tup{q',e'}\in R_\mathrm{chm}$ where $e'$ is obtained from $e$ by adding $o$ with an appropriate value of $\mathit{inv}(o)$ and an empty visibility.

  The transition corresponding to the atomic section of \texttt{put} is labeled by a sequence of visibility actions (one for each linearized operation) followed by a linearization action. Let $\sigma$ denote this sequence of actions. This transition leads to a state $q'$ where the array \texttt{table} may have changed (unless writing the same value), and the history variable \texttt{lin} is extended with the \texttt{put} operation $o$ executing this step. We define an abstract execution $e'$ from $e$ by changing $\mathit{lin}$ to the new value of \texttt{lin}, and defining an absolute visibility for $o$. We have that $e \xrightarrow{\sigma} e'$ because $e'$ is consistent with $W_\mathrm{m}$. Also, $\tup{q',e'}\in R_\mathrm{chm}$ because the validity of (\ref{it:unspec1}), (\ref{it:unspec2}), and (\ref{it:spec1}) follow directly from the definition of $e'$. The atomic section of \texttt{get} can be handled in a similar way. The simulation of return actions of \texttt{get} operations is a direct consequence of point (\ref{it:spec2}) which ensures consistency with $S$.

  For \texttt{has}, we focus on the atomic sections containing \texttt{vis} commands and the linearization commands (the other internal steps are simulated by $\epsilon$ steps of $\denot{W_\mathrm{m}}_\mathrm{s}$, and the simulation of the return step follows directly from (\ref{it:spec3}) which justifies the consistency of the return value). The atomic section around the procedure call corresponds to a transition labeled by a sequence $\sigma$ of visibility actions (one for each linearized modifier) and leads to a state $q'$ with a new active \texttt{has} operation $o$ (compared to $q$). We have that $e \xrightarrow{\sigma} e'$ because $e'$ is consistent with $W_\mathrm{m}$. Indeed, the visibility of $o$ in $e'$ is not constrained since $o$ has not been linearized and the $W_\mathrm{m}$-consistency of $e'$ follows from the $W_\mathrm{m}$-consistency of $e$. Also, $\tup{q',e'}\in R_\mathrm{chm}$ because $\mathit{index}(o)=-1$ and (\ref{it:spec3}) is clearly valid. The atomic section around the read of \texttt{table[k]} is simulated by $\denot{W_\mathrm{m}}_\mathrm{s}$ in a similar way, noticing that (\ref{it:spec3}) models precisely the effect of the visibility commands inside this atomic section. For the simulation of the linearization commands is important to notice that any active \texttt{has} operation in $e$ has a visibility that contains all modifiers which returned before it was called and as explained above, this visibility is monotonic.
\end{proof}

\end{document}